\documentclass[aps,prd,twocolumn,showpacs,floatfix,preprintnumbers,amsfont,amsmath,amssymb,nofootinbib, superscriptaddress]{revtex4}

\usepackage{color}
\usepackage[font=small]{caption,subfig}
\usepackage{hyperref}
\usepackage[normalem]{ulem}
\usepackage{lipsum}
\usepackage{url}
\usepackage{float}
\usepackage{ctable}
\usepackage{graphicx}
\usepackage[font=small,
   justification=justified,
   format=plain]{caption} 

\newcommand{\sk}[1]{}

\newcommand{\reffig}[1]{Figure~\ref{fig:#1}}

\newcommand{\be}{\begin{equation}}
\newcommand{\ee}{\end{equation}}
\newcommand{\ba}{\begin{eqnarray}}
\newcommand{\ea}{\end{eqnarray}}

\newcommand{\aap}{Astron. Astrophys.}

\newcommand{\pasp}{Publ. Astron. Soc. Pac.}

\allowdisplaybreaks 

\begin{document}

\title{Parameter Estimation for GW170817 using Relative Binning}

\author{Liang Dai}
\email{ldai@ias.edu}
\thanks{NASA Einstein Fellow}
\affiliation{\mbox{School of Natural Sciences, Institute for Advanced Study, 1 Einstein Drive, Princeton, New Jersey 08540, USA}}
\author{Tejaswi Venumadhav} 
\affiliation{\mbox{School of Natural Sciences, Institute for Advanced Study, 1 Einstein Drive, Princeton, New Jersey 08540, USA}}
\author{Barak Zackay}
\affiliation{\mbox{School of Natural Sciences, Institute for Advanced Study, 1 Einstein Drive, Princeton, New Jersey 08540, USA}}

\date{\today}


\begin{abstract}

Relative binning is a new method for fast and accurate evaluation of the likelihood of gravitational wave strain data. This technique can be used to produce reliable posterior distributions for compact object mergers with very moderate computational resources. We use a fast likelihood evaluation code based on this technique to estimate the parameters of the double neutron-star merger event GW170817 using publicly available LIGO data. We obtain statistically similar posteriors using either Markov-chain Monte-Carlo or nested sampling. The results do not favor non-zero aligned spins at a statistically significant level. There is no significant sign of non-zero tidal deformability (as quantified by the Bayesian evidence), whether or not high-spin or low-spin priors are adopted. Our posterior samples are publicly available, and we also provide a tutorial \texttt{Python} code to implement fast likelihood evaluation using the relative binning method.

\end{abstract}


\maketitle

Relative binning is a simple but powerful technique to speed up likelihood evaluations in gravitational wave parameter estimation~\cite{binning}. In this research note, we use the relative binning method to analyze the binary neutron-star merger event GW170817~\cite{abbott2017gw170817} using publicly available LIGO data~\cite{Vallisneri:2014vxa}. This event lasted for about $\sim 4000$ wave cycles (from $f = 23\,$Hz) before it exited the LIGO band; of all the compact object mergers detected so far, GW170817 covered the widest range of detectable frequencies and thus requires the finest search in parameter space. It is computationally challenging to perform parameter estimation for this merger using conventional methods (in contrast to the binary black hole mergers detected previously). 

We analyze the full $2048$ second-long sample of strain data from each LIGO detector that was released (at a sampling rate of $4096\,$Hz). We estimate the power spectrum density (PSD) from the noise-subtracted data itself. For a chirp mass $\mathcal{M}^{\rm det}$ in the detector frame and a symmetric mass ratio $\eta$, the gravitational wave frequency at the last stable circular orbit is given by $f_{\rm ISCO} \approx 1600\,{\rm Hz}\,(\mathcal{M}^{\rm det}/1.1976\,M_\odot)^{-1}\,(\eta/0.25)^{0.6}$ assuming a non-spinning binary. Since the waveform after the circular inspiral phase is theoretically uncertain for neutron-star mergers, we restrict to the frequency range $23\,{\rm Hz} < f < 1500\,{\rm Hz}$ when computing the likelihood in the frequency domain.

In order to easily compare our results to those in Ref.~\cite{Abbott:2018wiz}, we adopt the phenomenological frequency-domain waveform model \texttt{IMRPhenomD\_NRTidal} as implemented in the publicly available \texttt{LALSuite} library. This model was obtained by augmenting the binary-black-hole waveform model \texttt{IMRPhenomD}~\cite{Husa:2015iqa,Khan:2015jqa} with the numerically calibrated effects of tidal deformability~\cite{Dietrich:2017aum, Dietrich:2018uni}.

We allow six intrinsic parameters for the merger: detector-frame chirp mass $\mathcal{M}^{\rm det}$, symmetric mass ratio $\eta$, spin components along the direction of orbital angular momentum $\chi_{1z}$ and $\chi_{2z}$, and tidal deformation parameters $\Lambda_1$ and $\Lambda_2$~\cite{favata2014systematic}. We neglect the phasing effects of in-plane spins. In addition, for each LIGO detector we consider a set of extrinsic parameters---an effective distance $D_{\rm eff}$, a phase constant $\phi_c$, and an arrival time $t_c$---a total of six extrinsic parameters. We reduce the effective dimension of the parameter space by treating the two sets of extrinsic parameters independently, and separately (and analytically) maximizing the likelihood with respect to the two $D_{\rm eff}$'s and the two $\phi_c$'s. This simplification increases the model degrees of freedom by neglecting amplitude, phase and time correlations between the detectors. The amplitude parameters are poorly determined in any case, so this should not lead to parameter biases and should only slightly worsen parameter uncertainties~\cite{RZ}.

We adopt the same prior for intrinsic parameters as that used in Ref.\cite{Abbott:2018wiz}. In this prior, the two component masses are drawn from flat distributions in the range $[0.5,\,7.7]\,M_\odot$. We adopt the ``high-spin'' prior for the spins, in which the moduli and the directions of the two (dimensionless) spin vectors, $\boldsymbol{\chi}_i,\,i=1,2$, are drawn uniformly within the range $[0, 0.89]$, and randomly on the unit sphere, respectively. We generate spin vectors, and then pass their aligned components $\chi_{1z}$ and $\chi_{2z}$ to the waveform generator. Furthermore, we impose flat priors separately on the two tidal deformation parameters $\Lambda_1,\,\Lambda_2 \in [0,\,5000]$. For the two arrival times $t_{c,1}$ and $t_{c, 2}$, we allow a window $[-0.005,\,0.005]\,$sec around the most probable value for either of them.

To utilize relative binning, we first compute summary data using a fiducial waveform $h_0$ that corresponds to crudely-estimated parameter values taken from Ref.~\cite{Abbott:2018wiz}. We then refine the values of the best-fit parameters by maximizing the likelihood (computed using the summary data as in Ref.~\cite{binning}). We then use these improved parameter values, which are close but not necessarily equal to the peak of the likelihood, to update the fiducial waveform and the summary data. In principle, we can iterate this step several times, but in practice, the accuracy of the likelihood evaluation routine is already sufficiently good for parameter estimation in the vicinity of the best-fit solution. With about $60$ frequency bins, we find that the absolute error on the log-likelihood function $\Delta\ln\mathcal{L}$ is controlled to be $\Delta\ln\mathcal{L} < 0.01\,(\ln\mathcal{L}_{\rm max} - \ln\mathcal{L})$~\cite{binning}.

In order to carry out parameter inference, we couple our likelihood evaluation routine to the Markov-chain Monte-Carlo sampler \texttt{emcee}~\cite{goodman2010ensemble, 2013PASP..125..306F}. To speed up convergence, we restrict the search to be within the region $\mathcal{M}^{\rm det} \in [1.1972,\,1.1982]\,M_\odot$ and $\eta \in [0.15,\,0.25]$.

\begin{figure*}[t]
\begin{center}
  \includegraphics[scale=0.45]{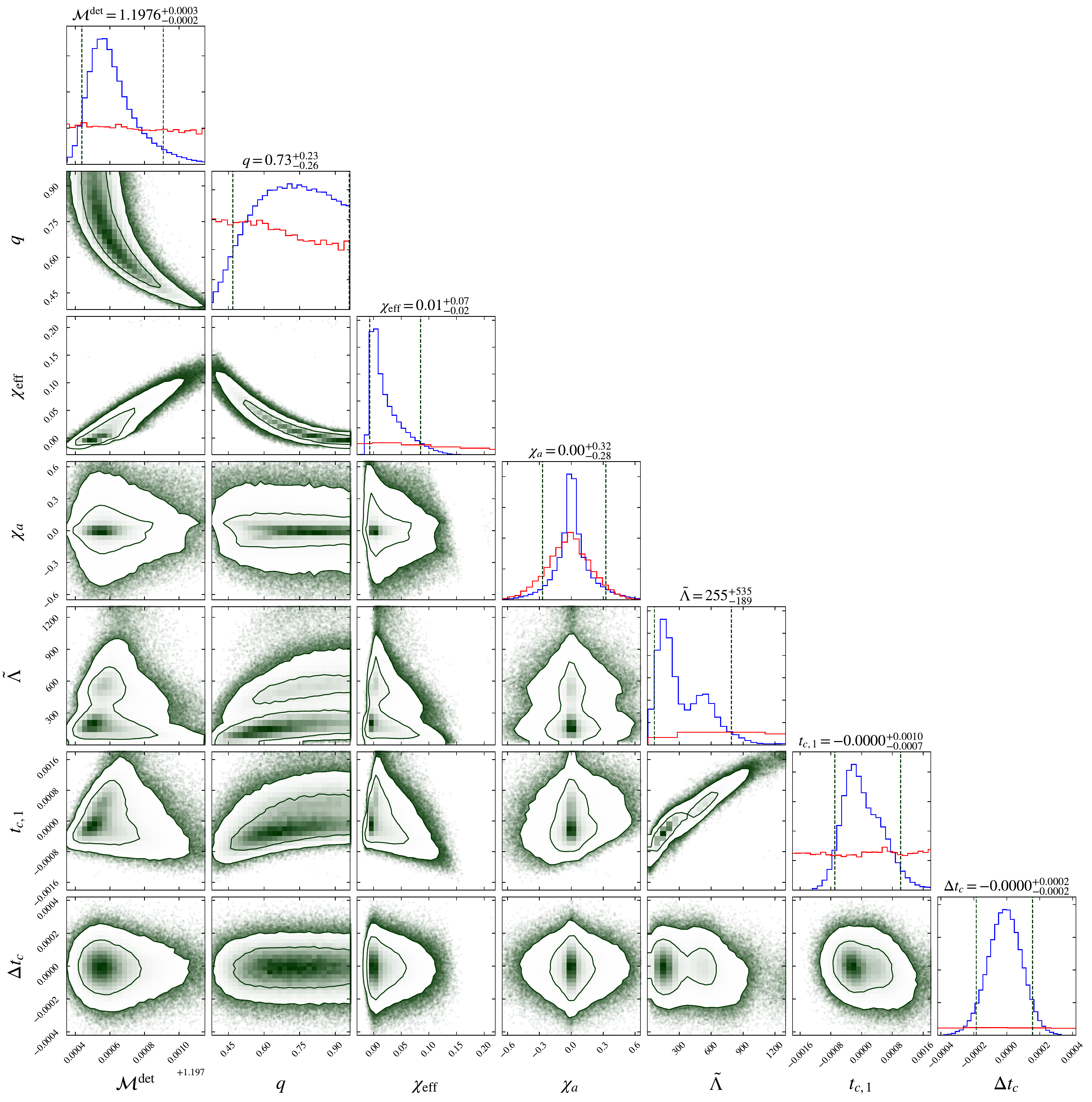}
  \caption{\label{fig:post_emcee_fmax1D5kHz} Posterior distributions obtained using the \texttt{emcee} sampler. Red and blue histograms in the diagonal plots show the prior and marginalized posterior distributions, respectively. The dashed vertical lines and labels show the $5\%$ and $95\%$ quantiles for the parameters. Off-diagonal plots show the two-dimensional joint posterior distributions, with contours corresponding to the $68\%$ and $95\%$ quantiles. The parameters shown are the detector-frame chirp mass $\mathcal{M}^{\rm det} = (1+z)\,\mathcal{M}\,$ ($M_\odot$), mass ratio $q$, effective spin parameter $\chi_{\rm eff}=(m_1\,\chi_{1z} + m_2\,\chi_{2z})/(m_1 + m_2)$ and anti-symmetric combination $\chi_a = (\chi_{1z} - \chi_{2z})/2$ built from the aligned spins, reduced tidal deformability parameter $\tilde\Lambda$, the arrival time at Livingston $t_{c,1}$, and the difference in arrival times between Hanford and Livingston $\Delta t_c = t_{c, 2} - t_{c, 1}$. The zero points for $t_{c, 1}$ and $t_{c, 2}$ are separately chosen, so their absolute values carry no physical meaning. The \texttt{IMRPhenomD\_NRTidal} waveform model was used for evaluating the likelihood.}
\end{center}
\end{figure*}

We collected a total of $\simeq 3\times 10^8$ parameter samples (among which $\sim 10^5$--$10^6$ are independent) within $\simeq 150$ core hours. \reffig{post_emcee_fmax1D5kHz} shows the posterior distributions using a thinned set of $\simeq 3\times 10^5$ samples. In general, our results are in good agreement with those of Ref.~\cite{Abbott:2018wiz}. Points of agreement include the severe $(q,\,\chi_{\rm eff})$ degeneracy (where $\chi_{\rm eff} = (m_1 \chi_{1, z} + m_2 \chi_{2, z})/(m_1 + m_2)$ is the ``effective spin parameter''), the skewed distribution for $\chi_{\rm eff}$, and a marginal hint of non-zero tidal deformability. 

\begin{figure*}[t]
\begin{center}
  \includegraphics[scale=0.45]{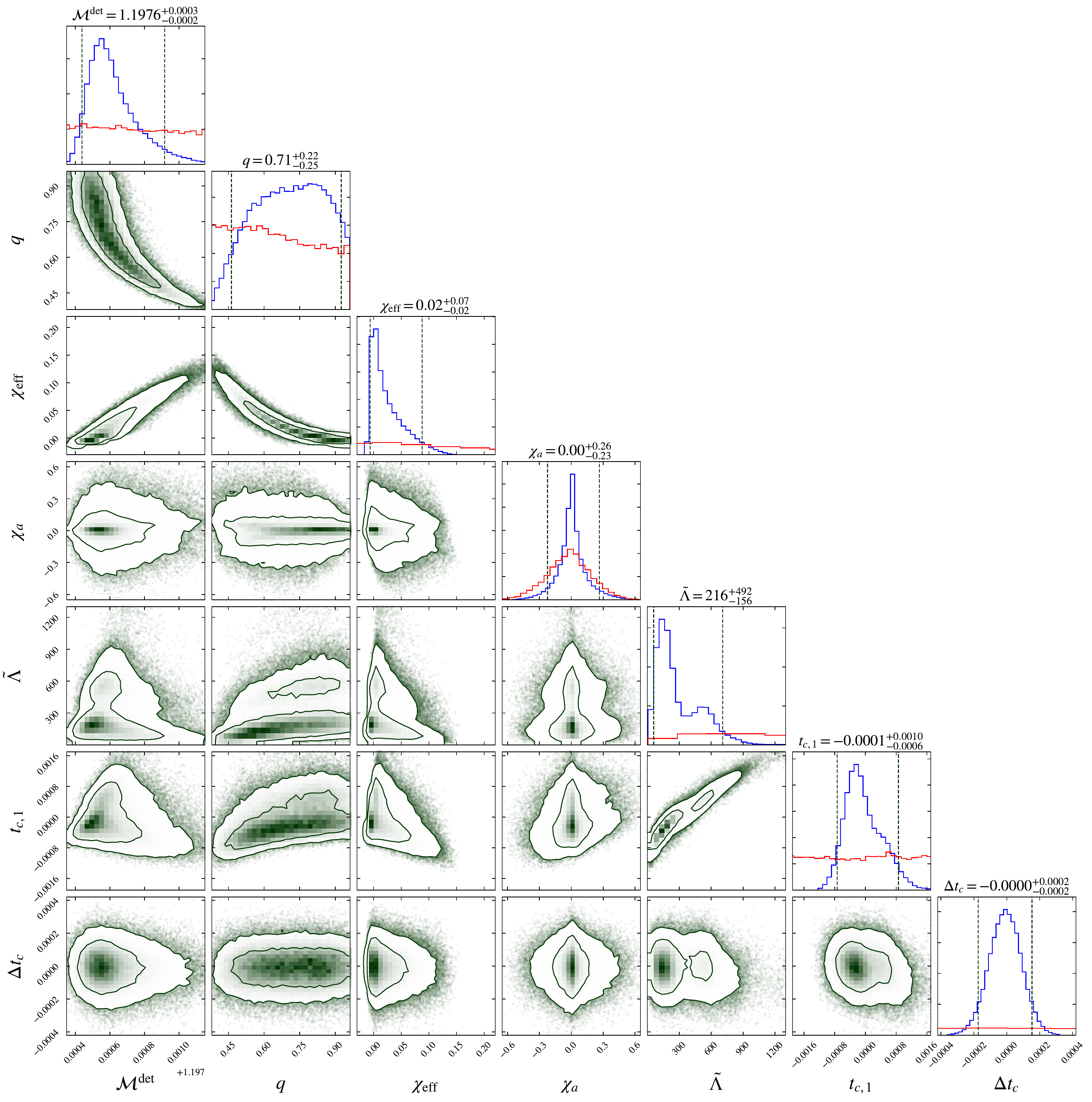}
  \caption{\label{fig:post_multinest_fmax1D5kHz} The posterior distributions obtained using the \texttt{MultiNest} sampler. The \texttt{IMRPhenomD\_NRTidal} waveform model is used for evaluating the likelihood. Frequencies $23\,{\rm Hz} < f < 1500\,{\rm Hz}$ are included in the analysis.}
\end{center}
\end{figure*}

\begin{figure*}[t]
\begin{center}
  \includegraphics[scale=0.45]{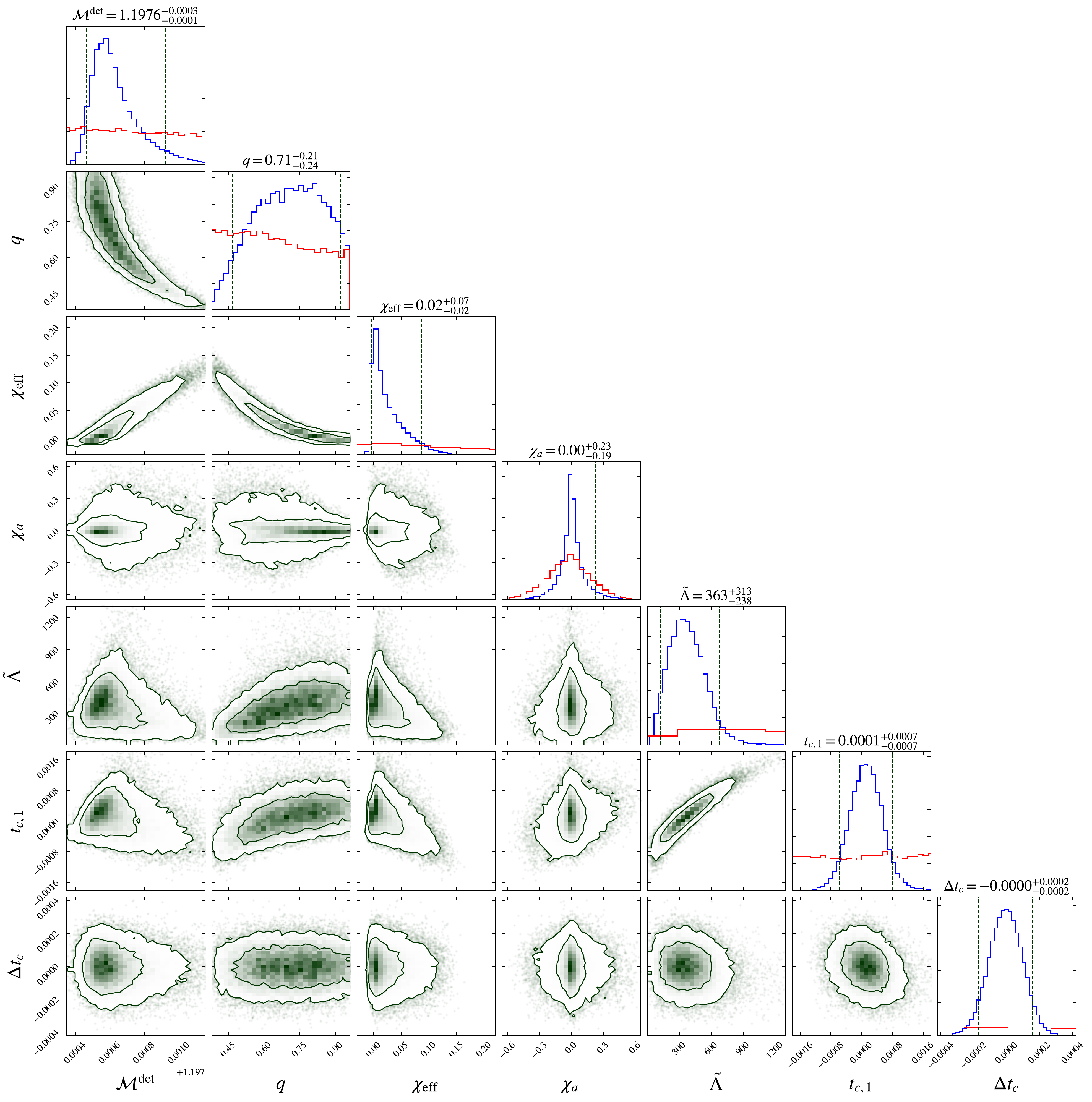}
  \caption{\label{fig:post_multinest_fmax1kHz} The posterior distributions obtained using the \texttt{MultiNest} sampler. The \texttt{IMRPhenomD\_NRTidal} waveform model is used to evaluate the likelihood. Unlike \reffig{post_multinest_fmax1D5kHz}, frequencies $23\,{\rm Hz} < f < 1000\,{\rm Hz}$ are included in the analysis.}
\end{center}
\end{figure*}

When including frequencies up to $1500\,{\rm Hz}$, we obtain a bimodal posterior distribution for the reduced tidal deformability parameter $\tilde\Lambda$ (c.f. Eq.(5) of Ref.\cite{Abbott:2018wiz}), with a major peak at $\tilde\Lambda\simeq 200$ and a minor peak at $\tilde\Lambda\simeq 500$ (see \reffig{post_emcee_fmax1D5kHz} or \reffig{post_multinest_fmax1D5kHz}). This is in good agreement with the results shown in Ref.\cite{Abbott:2018wiz}. \reffig{post_multinest_fmax1kHz} shows, however, that the posterior distribution for $\tilde\Lambda$ becomes singly peaked at $\tilde\Lambda \simeq 400$ if only frequencies $f < 1000\,{\rm Hz}$ are included in the analysis. 

We expect to see changes to the posteriors for $\tilde{\Lambda}$ after including higher frequencies, since the correction to the binary orbital phase due to tidal deformation rapidly increases with frequency. However, we should exercise caution in interpreting any features in the posteriors for GW170817 due to data above $1000\,{\rm Hz}$, since there is negligible signal-to-noise in this range of frequencies due to the rapid rise in the detector noise (in terms of the log-likelihood, $\Delta \ln{\mathcal{L}} \sim 0.5$). This suggests that the amount of information is insufficient to distinguish between random detector noise and changes to the physical signal due to tidal deformation; later in this note, we quantify this by calculating the Bayesian evidence.

A few other notable features of the posterior distributions are as follows:
\begin{enumerate}

  \item The well known degeneracy between $q$ and $\chi_{\rm eff}$ is actually a tri-variate degeneracy involving the chirp mass $\mathcal{M}^{\rm det}$. 
  
  \item The effective spin for the aligned spins is constrained to be low ($-0.02 < \chi_{\rm eff} < 0.07$ at the $90 \%$ level), but the anti-symmetric term, $\chi_a = (\chi_{1, z} - \chi_{2, z})/2$, is barely constrained by the data. 
  
  \item There is a significant degeneracy between $\tilde\Lambda$ and the common merger times at the two detectors. In general, parameters that occur with positive powers of frequency in the post-Newtonian expansion of the gravitational wave phase can be degenerate with each other (and similarly for the ones with negative powers). When frequencies up to $1500\,{\rm Hz}$ are included in the analysis, the two peaks of the posterior distribution of $\tilde\Lambda$ also correspond to two different values for the common arrival time (as can be seen from \reffig{post_emcee_fmax1D5kHz} or \reffig{post_multinest_fmax1D5kHz}).
  
  \item Finally, the arrival time difference between the two LIGO sites, $\Delta t_c = t_{c, 2} - t_{c, 1}$, is statistically independent of any of the intrinsic parameters (not perfectly so with $\tilde\Lambda$ though). This is consistent with the fact that the strain signals at both detectors originated from a common source on the sky. This also justifies our approximation that the amplitudes, phases and the arrival times at both detectors can be treated as independent parameters. 
  
\end{enumerate}

We also checked that the posterior distributions are unchanged when the number of frequency bins is increased from $\simeq 60$ to $\simeq 100$.

As a final check, we also used \texttt{pyMultiNest}~\cite{2014A&A...564A.125B} to estimate the posterior distributions. The \texttt{pyMultiNest} code implements the technique of multi-modal nested sampling; as the name suggests, it is especially advantageous when the posterior distribution has multiple modes. We obtain $\simeq 10^6$ samples using even less computational resources than in the previous case. \reffig{post_multinest_fmax1D5kHz} shows the resulting posterior distributions; they are in excellent agreement with the \texttt{emcee} results.

\begin{figure*}[t]
\begin{center}
  \includegraphics[scale=0.45]{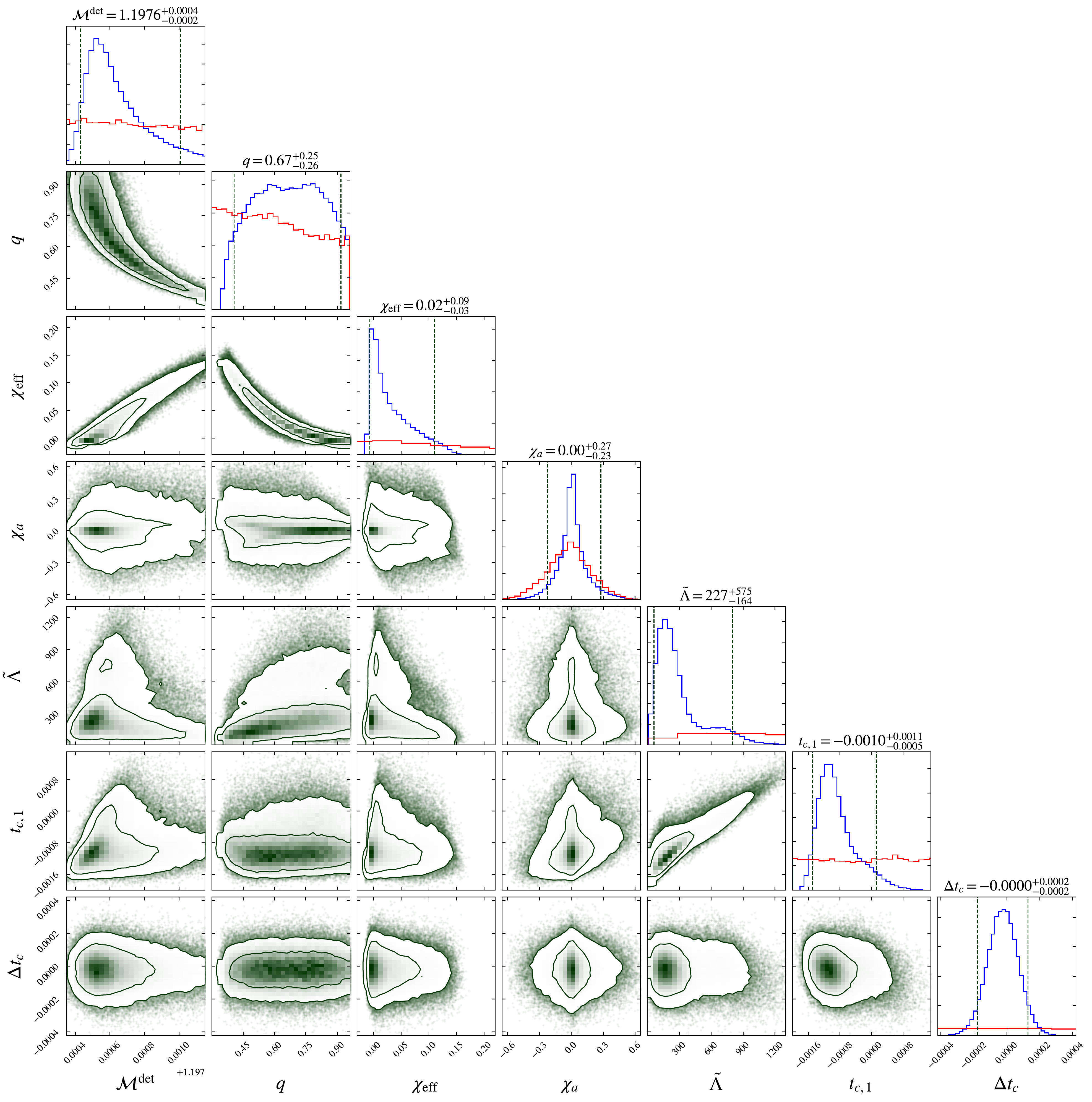}
  \caption{\label{fig:post_multinest_TaylorF2_fmax1D5kHz} The posterior distributions obtained using the \texttt{MultiNest} sampler. The analytic \texttt{TaylorF2} waveform model is used to evaluate the likelihood. Frequencies $23\,{\rm Hz} < f < 1500\,{\rm Hz}$ are included in the analysis.}
\end{center}
\end{figure*}

\begin{figure*}[t]
\begin{center}
  \includegraphics[scale=0.45]{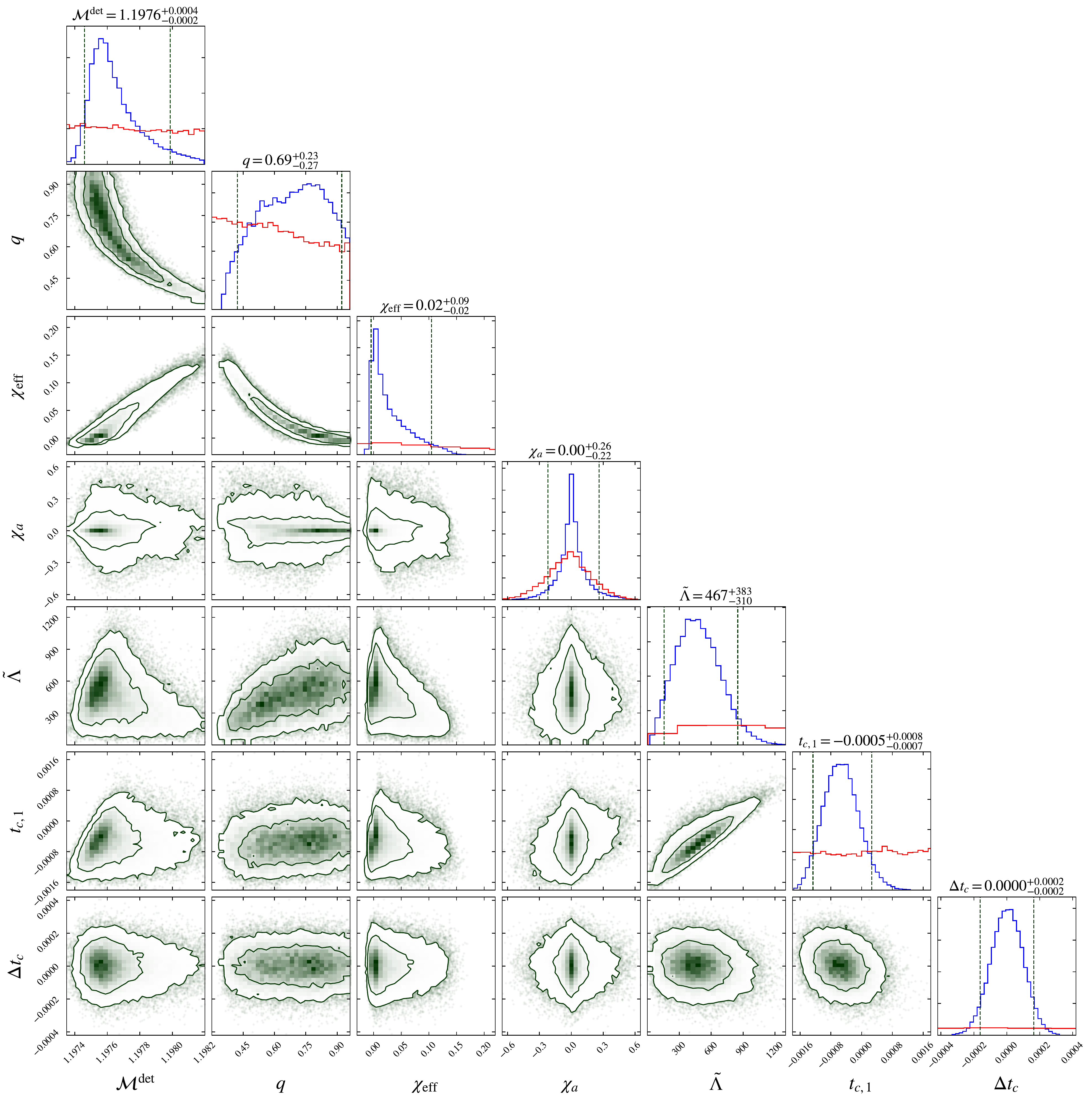}
  \caption{\label{fig:post_multinest_TaylorF2_fmax1kHz} The posterior distributions obtained using the \texttt{MultiNest} sampler. The analytic \texttt{TaylorF2} waveform model is used to evaluate the likelihood. Unlike \reffig{post_multinest_TaylorF2_fmax1D5kHz}, frequencies $23\,{\rm Hz} < f < 1000\,{\rm Hz}$ are included in the analysis.}
\end{center}
\end{figure*}

The most significant benefit of a fast parameter estimation pipeline is the ability to quickly test models using gravitational wave data. We show a few simple examples of this. One basic test one might imagine performing is to compare different waveform models. \reffig{post_multinest_TaylorF2_fmax1D5kHz} and \reffig{post_multinest_TaylorF2_fmax1kHz} show the posteriors when the likelihood is evaluated using the \texttt{TaylorF2} waveform model, for $f < 1500\,{\rm Hz}$ and $f < 1000\,{\rm Hz}$, respectively. These two figures are similar to \reffig{post_multinest_fmax1D5kHz} and \reffig{post_multinest_fmax1kHz} respectively in other aspects. The analytic model \texttt{TaylorF2} uses the stationary phase approximation for the waveform, and uses the 3.5PN expression for the orbital phase of inspiraling binary black holes with aligned spins~\cite{sathyaprakash1991choice, Bohe:2013cla, Arun:2008kb, Mikoczi:2005dn, Bohe:2015ana, Mishra:2016whh}, along with the tidal effects on the phase (up to the 6PN level) for compact stars~\cite{PhysRevD.83.084051}. The posterior distributions for the \texttt{TaylorF2} case are largely compatible with, but not identical to those obtained using the \texttt{IMRPhenomD\_NRTidal} waveform model. Although in principle \texttt{IMRPhenomD\_NRTidal} captures the later stage of the inspiral better than \texttt{TaylorF2} does, the differences in the best-fit values of the intrinsic parameters are not statistically significant. We note that in the case with a maximum allowed frequency of $1500\,{\rm Hz}$, the secondary bump in the posterior distribution of $\tilde\Lambda$ weakens substantially compared to the results using \texttt{IMRPhenomD\_NRTidal},  which was also found in Ref.~\cite{Abbott:2018wiz}.

\begin{table}[!t]
\centering
\begin{tabular}{ |c|c| }
\hline
  Allowed parameters & $ \ln Z$ \\ 
 \hline
 \hline
 All parameters & $554.82 \pm 0.06$  \\
 \hline 
  Equal masses & $554.59 \pm 0.06$ \\
 \hline 
 No spins & $556.41 \pm 0.06$ \\
 \hline 
 No tides & $557.87 \pm 0.05$ \\
 \hline
 No spins or tides & $557.61 \pm 0.05$ \\
 \hline
 Equal masses, no spins or tides & $549.12\pm 0.04$ \\
 \hline
\end{tabular}
\caption{\label{tab:lnZ_fmax1D5kHz} Relative log-Bayesian evidences computed for restricted \texttt{IMRPhenomD\_NRTidal} waveform models. Frequencies $23\,{\rm Hz} < f < 1500\,{\rm Hz}$ are used. Note that the zero-point of $\ln Z$ is arbitrary.}
\end{table}

Another test that can be performed is to check whether a set of physical parameters are needed to fit the data. We can naturally accomplish this by comparing the Bayesian evidences between models with and without these parameters. Table \ref{tab:lnZ_fmax1D5kHz} shows the relative log-Bayesian evidences as computed by \texttt{pyMultiNest} using the \texttt{IMPhenomD\_NRTidal} waveform, for a few cases in which some of the intrinsic parameters of the mergers were held fixed. According to commonly adopted standards of Bayesian model selection, the changes in the log-Bayesian evidence between different cases are not large enough to be considered strongly discriminative. The somewhat extreme case with mass asymmetry, spins and tidal deformability all disallowed, however, is clearly disfavored. Subject to our prior assumptions, the results show no significant evidence for non-zero aligned spins of the binary components. Also, regardless of the assumption of either high or low spin magnitudes, there is no significant evidence for non-zero tidal deformability.

Other inferences using the data can be easily performed given the full likelihood, which could either be efficiently re-computed using relative binning, or be accessed using samples from the posterior distribution along with a known and sufficiently well-behaved prior. To this end, we make both our posterior samples, as well as a sample \texttt{Python} code for implementing fast likelihood evaluation, publicly available at \url{https://bitbucket.org/dailiang8/gwbinning/}.

\textit{Acknowledgments:} 
This research has made use of data, software and/or web tools obtained from the LIGO Open Science Center (\url{https://losc.ligo.org}), a service of LIGO Laboratory, the LIGO Scientific Collaboration and the Virgo Collaboration. LIGO is funded by the U.S. National Science Foundation. Virgo is funded by the French Centre National de Recherche Scientifique (CNRS), the Italian Istituto Nazionale della Fisica Nucleare (INFN) and the Dutch Nikhef, with contributions by Polish and Hungarian institutes.

We thank Vera Gluscevic, Soichiro Morisaki, David Radice, Javier Roulet, Hideyuki Tagoshi, and Matias Zaldarriaga for helpful discussions. LD is supported at the Institute for Advanced Study by NASA through Einstein Postdoctoral Fellowship grant number PF5-160135 awarded by the Chandra X-ray Center, which is operated by the Smithsonian Astrophysical Observatory for NASA under contract NAS8-03060. TV acknowledges support from the Schmidt Fellowship and the W.M. Keck Foundation Fund. BZ acknowledges support from the Infosys Membership Fund.

\bibliographystyle{apsrev4-1-etal}
\bibliography{gw}

\clearpage

\end{document}